\documentclass[twocolumn,floatfix]{revtex4}

\usepackage{graphicx}
\usepackage{dcolumn}
\usepackage{amsmath}
\usepackage{amsmath,amsfonts,amssymb}
\usepackage{wrapfig}
\usepackage{graphicx}
\usepackage{bbm}
\usepackage{graphics}

\setcitestyle{super}

\def \beq {\begin{equation}}
\def \eeq {\end{equation}}
\def \ba {\begin{eqnarray}}
\def \ea {\end{eqnarray}}

\begin{document}

\hsize\textwidth\columnwidth\hsize\csname@twocolumnfalse
\endcsname

\title{Modular Cryogenic Interconnects for Multi-Qubit Devices}
\author{J. I. Colless and D. J. Reilly$^*$}
\affiliation{ARC Centre of Excellence for Engineered Quantum Systems, School of Physics, The University of Sydney, Sydney, NSW 2006, Australia}

\begin{abstract}
We have developed a modular interconnect platform for the control and readout of multiple solid-state qubits at cryogenic temperatures. The setup provides 74 filtered dc-bias connections, 32 control and readout connections with a bandwidth above 5 GHz, and 4 microwave feed lines that allow operation to above 10 GHz. The incorporation of a radio-frequency (rf) interposer enables the platform to be separated into two printed circuit boards, decoupling the simple board that is bonded to the qubit chip from the multilayer board that incorporates expensive connectors and components. This modular approach lifts the burden of duplicating complex interconnect circuits for every prototype device. We report the performance of this platform at milli-Kelvin temperatures, including signal transmission and crosstalk measurements.
\end{abstract}

\maketitle
\section{Introduction}
Computationally-useful quantum machines will likely require the manipulation and readout of millions of physical qubits \cite{Suchara}, controllably interacting with each other \cite{yale,martinis,yacoby}, and connected to a complex layer of classical hardware \cite{Hornibrook}. Analogous to the operation of modern integrated circuits, its conceivable that in the future the density of connections required for quantum computation can be achieved by lithographically integrating qubits into sub-systems on-chip, such that it is no longer necessary to directly address each individual component of the circuit from the outside. Today however, demonstrating and debugging the operation of multi-qubit technologies requires direct access to every aspect of the device, typically controlled by external circuits beyond the qubit chip. For most prototype experiments this means high-bandwidth, high-density wiring, connecting room temperature electronics to quantum systems at the base temperature of a dilution refrigerator.

The use of printed circuit boards (PCBs) and miniaturised high-frequency connectors \cite{Colless:2011} has become the standard interface between cryogenic cabling and the qubit device chip. Connectors bring the signals onto the PCB and bond-wires connect tracks and transmission lines to bond-pads on the chip that comprises the quantum device. As circuits become more complex and the number of electrical connections grows however, producing a new PCB for every new chip is both costly and time consuming since de-bonding a functional device to reuse the PCB and connectors is impractical. 

Here we report a high wire-count, small-footprint interconnect-platform that functions at temperatures below 20 milli-Kelvin and alleviates the burden of duplicating the expensive and complex circuit boards for every functional device. The incorporation of a radio-frequency (rf) interposer enables the platform to be separated into two PCBs, decoupling the simple board that is bonded to the qubit chip from the complex multilayer board that incorporates connectors and components.  The `signal board' includes all the dc, radio-frequency and microwave signal interconnects, as well as filters and bias tees. The smaller and simpler `device board' implements only the bond pads to allow wire bonding to the chip. 
  This modular system has the following further advantages over traditional interconnect methods used for qubit devices\cite{Reilly:2007}:

$\bullet$ Custom device boards can be produced to allow chips of different sizes, different bonding arrangements, or multi-chip configurations to be accommodated without the need to redesign the interconnect circuitry. \par
$\bullet$ Chips can be permanently bonded to device boards for storage without the risk to the device from de-bonding in order to reuse the expensive PCB and connectors.\par
$\bullet$ Custom circuits can be included on the device board such as additional filtering or impedance matching networks, allowing the cryogenic circuitry to be easily extended as needed. \par
$\bullet$ With the device board interchangeable, the signal board and wiring can be fixed within a cryogenic system, reducing the likelihood of electrical faults which typically occur when making and breaking interconnects for measurement of a new device.\\

Below we demonstrate operation of this approach at cryogenic temperatures and present measurements characterising crosstalk and signal fidelity. While developed in the context of quantum dot based spin qubits, we anticipate that such a setup will be of general interest for the rapid-prototyping of nanoscale cryogenic circuits.
  
\begin{figure*}
\centering
\includegraphics[scale=0.5]{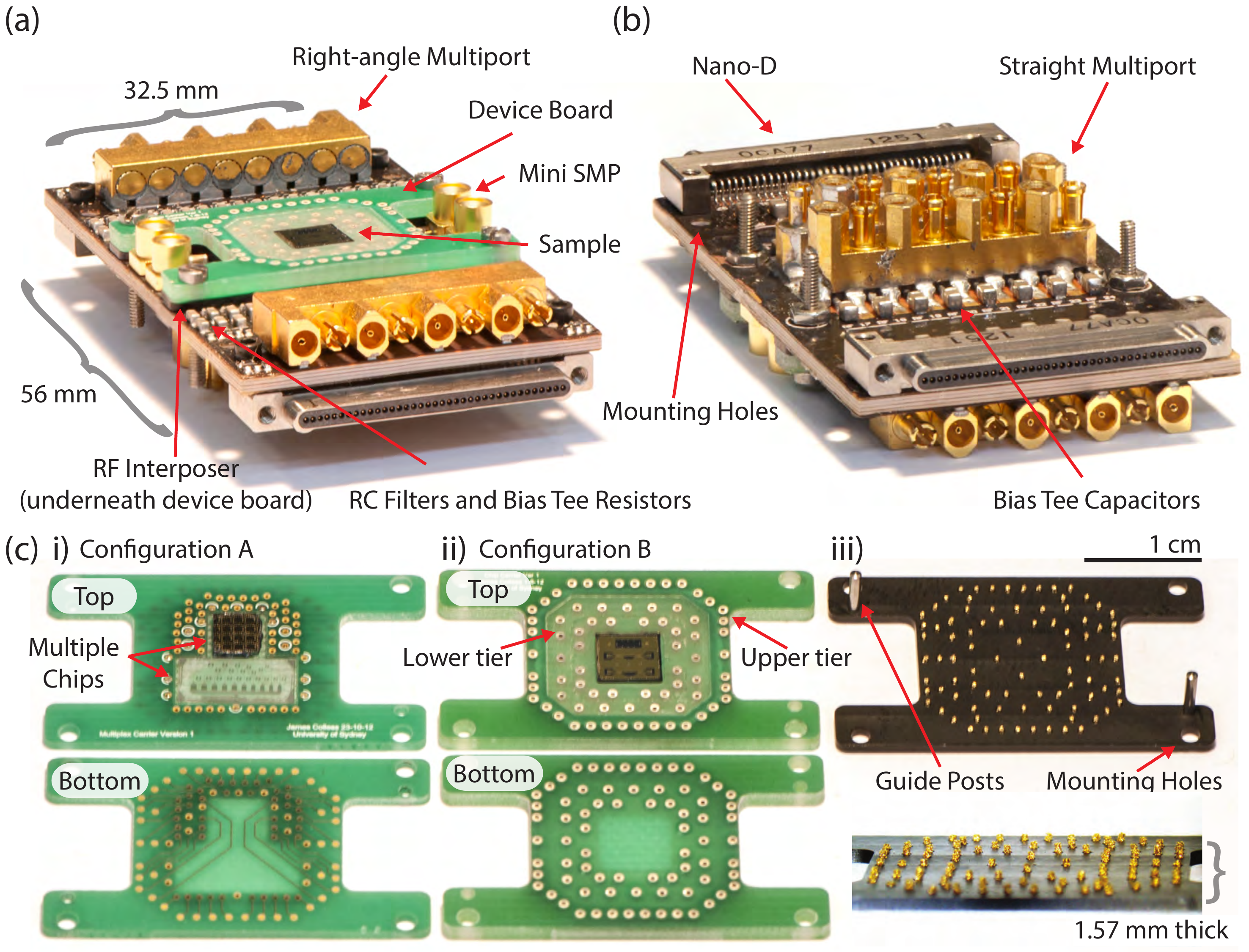}
\caption{\textbf{(a)} Top and \textbf{(b)} bottom isometric views of the assembled setup. \textbf{c) i)}  Top and bottom view of a device board configured for a multi-chip experiment where, in this design, bond-pad locations have been chosen to ensure easy wire bonding. On the bottom of the device board the copper pattern is identical to that on the top of the signal board (some interconnects are not used) with signal lines being routed on the surface of the signal board to their final location. \textbf{ii)} Top and bottom view of a device board configured for a single chip experiment requiring a large number of signal interconnects. In this case a two tiered structure is used to allow for optimal bonding, maximising rf signal integrity via short bond wires (inputs on the lower tier, closer to the device) while minimising the chance of a short with the dc wires (upper tier). \textbf{(c) iii)} Top and side-on view of the rf interposer with 74 `Fuzz button' interconnects, 2 alignment posts and 4 mounting holes.}
\label{fig:pcb_photos}
\end{figure*}

\begin{figure*}[H!]
\centering
\includegraphics[scale=.34]{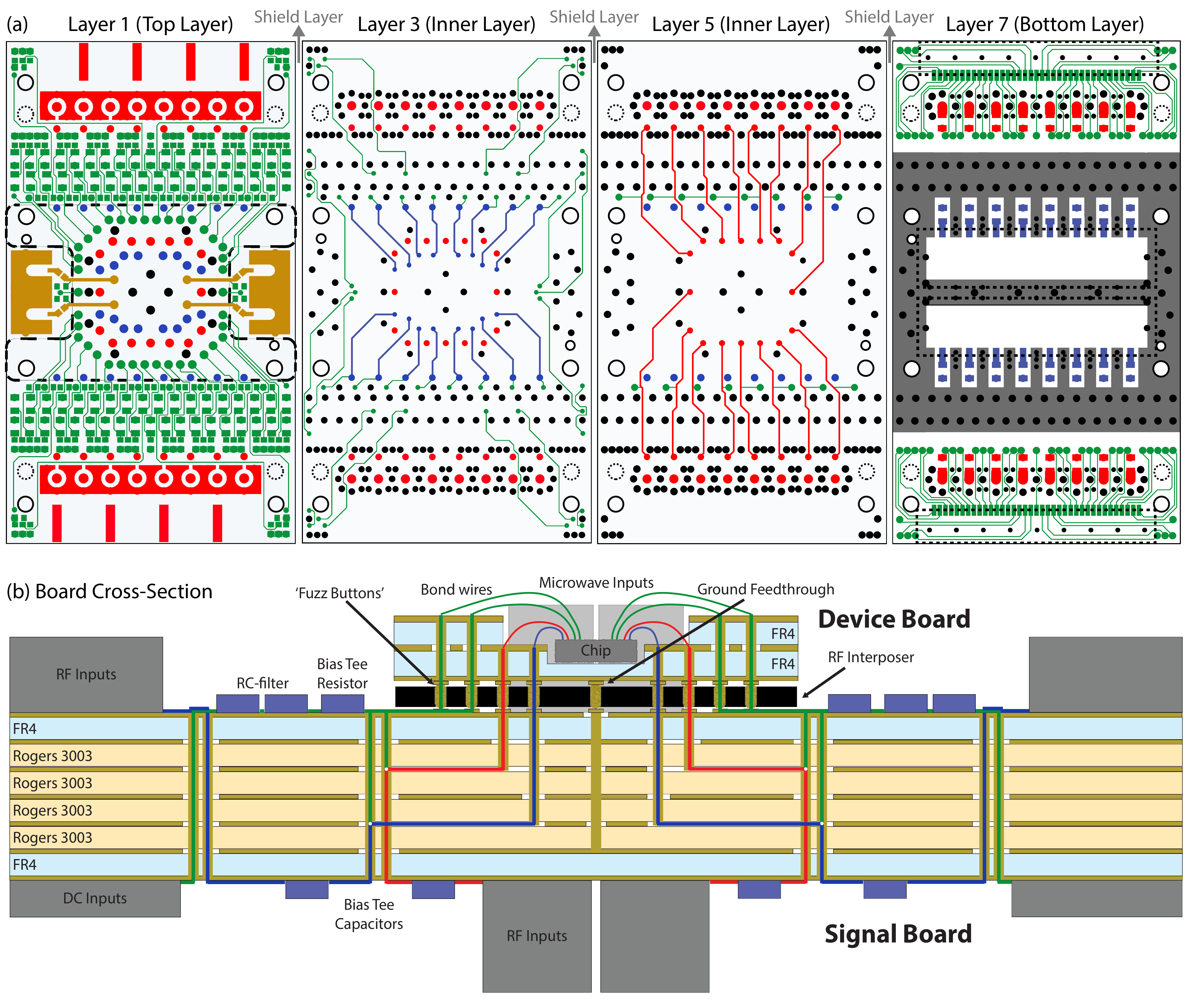}
\caption{\textbf{(a)} Signal board layers and {\textbf{(b)}} cross-section. \textbf{Layer 1 (top layer):} 4 microwave MSMP connectors (gold) are located on the top of the signal board along with tightly packed $RC$ filters and the dc biasing resistors for the bias tees (green). Right-angle Multiport rf connectors (red) launch 16 rf lines which are fed through to the bottom layer of the board. Holes near the corners of the board allow for mounting within a cryogenic system (dashed circles). The central part of the design contains the plated contacts for the rf interposer (interposer position shown with dashed black line) along with associated mounting holes and two guide holes (solid circles) to ensure correct alignment of the device board relative to the signal board. \textbf{Layer 3 (inner layer):} High-frequency signals launched from straight Multiport connectors get routed to their final location and pass through vias to their corresponding contact pad on the top of the board (blue). Bias lines fed via resistors on the top layer are added to each of the high-frequency lines (green). \textbf{Layer 5 (inner layer):} High-frequency signals launched from right-angle Multiport connectors get routed to their final location and pass through vias to their corresponding contact pad on the top of the board (red). Again, bias lines fed via resistors on the top layer are added to each of the high-frequency lines (green). \textbf{Layer 7 (bottom layer):} Straight Multiport rf connectors (dashed rectangles) launch 16 rf lines on the bottom layer of the board which, along with the 16 inputs from the top layer, pass through capacitors and into the middle of the stack (blue and red respectively). Twin 37-way nano-D connectors situated at either end of the board provide 74 dc inputs (green) which are then fed to the top layer of the board. Fencing vias and shielding ground planes are used throughout the design to prevent unwanted coupling, either between dc and rf layers or from one rf signal line to another.} 
\label{fig:pcb_layers}
\end{figure*}
\begin{figure*}[h!]
\centering
\includegraphics[scale=0.3]{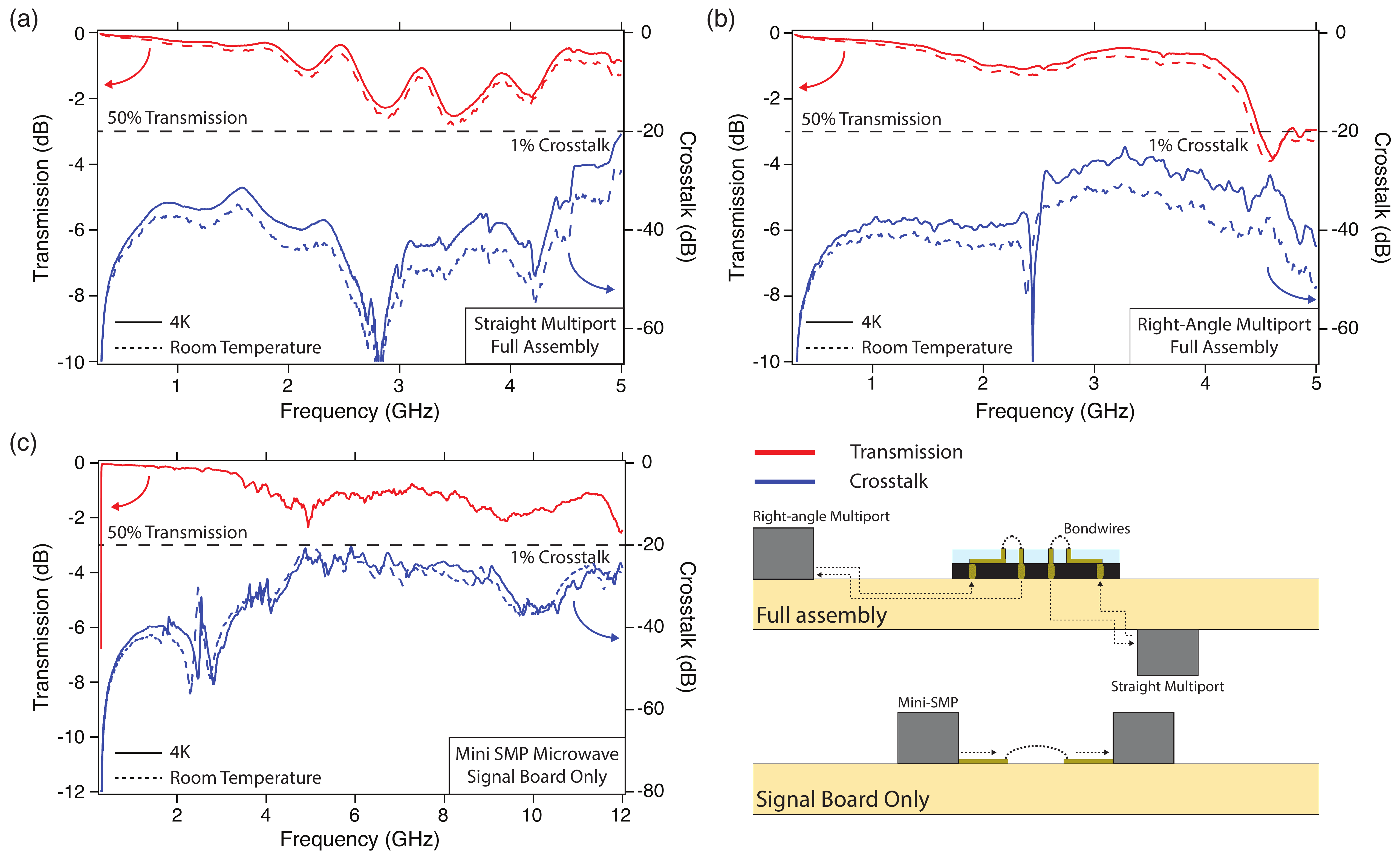}
\caption{\textbf{(a)} Transmission and crosstalk measurements for nearest neighbour paths launched from straight Multiport connectors which pass through the rf interposer and device board (see legend). Data at room temperature and at $4$ Kelvin, taken using a cryogenic probe station. \textbf{(b)} Transmission and crosstalk measurements for nearest neighbour paths launched from right-angle Multiport connectors which pass through the rf interposer and device board (see legend). \textbf{(c)} Transmission and crosstalk data for nearest neighbour microwave paths launched from mini-SMP connectors. Measurements are made directly on the signal board, without interposer and device board (see legend). The addition of the interposer and device board has little effect on the performance of the system below $5$ GHz.} 
\label{fig:measurements}
\end{figure*}

\section{System}
\subsection{Signal Board}
The signal PCB consists of a 32.5 mm $\times$ 56 mm, 7 layer stack, fabricated using a combination of Rogers 3003 laminate and standard FR4, bonded together with Arlon 6700 film. The lower dissipation factor ($\delta=0.0013$) of the Rogers 3003 at high frequency minimizes the loss of rf signals travelling through the inner layers of the board while the low dielectric constant ($\epsilon_{r}=3.00$) allows the ground planes to be moved closer to the signal lines, decreasing the amount of crosstalk between neighbouring tracks. Outer layers of FR4 dielectric increase the durability of the board and ensure more robust mounting of surface components. A total of 74 low-frequency bias inputs are provided by nano-D connectors\footnote[1]{Glenair Corp. Series 89 Nanominiature Connectors 890-013.} and are filtered via single-pole $RC$ filters ($R = 10$\ k$\Omega$ and $C = 15$\ nF, $f_{c} \sim 1$ kHz). Of this total, 36 dc lines then feed $RC$ bias tees ($R = 5$\ k$\Omega$ and $C = 47$\ nF, $f_{c} \sim 700$ Hz) where they are added to rf and microwave signal lines, while the 38 remaining dc lines proceed directly to the interposer. High-frequency rf inputs are coupled onto the board via high-density multiport connectors\footnote[2]{Rosenberger `Multiport' connectors, nominal maximum operating frequency of 18GHz.} which mount both normal and parallel to the PCB surface allowing 32 control and readout connections to the qubit chip. These rf signals travel on inner layers of the signal PCB where they are decoupled from both dc and microwave lines using ground planes and via fencing techniques to provide a quasi-coaxial geometry\cite{Colless:2011,viafence1,Shahparnia:2004ii}, as shown in Fig. 2(a). Finally, 4 microwave inputs are launched by Mini-SMP connectors\footnote[3]{Rosenberger MSMP, nominal maximum operating frequency of 65 GHz.} onto short transmission lines close to the interposer. Theses lines  are kept short in order to reduce loss and any parasitic coupling. All parts used on the board are nominally non-magnetic.

\subsection{Interposer}
The interposer is made of a thin plastic (Ultem 1000) 1.57 mm thick cut-out with 88 (74 signal and 14 ground) 0.38 mm holes drilled through the piece as shown in Fig. \ref{fig:pcb_photos}(c) iii). Electrical connections through these holes are made using captive `Fuzz Button' contact pins\footnote[4]{Custom Interconnects, Fuzz Buttons\textregistered}: tightly wound coils of Au/BeCu wire which act as a spring, pushing down against the signal board and up against the device board. These contacts are robust from room to milli-Kelvin temperatures and over several hundred mating cycles. If the contact pins are dislodged or damaged in handling they can be easily replaced by hand. Independent $S$-parameter measurements of an isolated interposer show that its intrinsic crosstalk and signal fidelity will not be a limiting factor in the performance of this platform up to 20 GHz. Guide posts are used to ensure correct alignment of the signal board/interposer/device board stack and mounting holes allow the entire assembly to be screwed  together firmly ensuring a robust electrical connection.

\subsection{Device Board}
Signals are routed from the signal board, through the interposer to the back face of the device board, where vias then bring them through to the top surface band-pads. The design of the device board can be tailored to meet experimental requirements so long as the outer shape of the board is no larger than the available space on the signal board (17.4 mm $\times$ 19.5 mm ignoring mounting and alignment offshoots). The location of the contact pads on the bottom of the device board must match that of the interposer itself (although contacts may be omitted if the full 88 signal and ground inputs are not required, as shown in Fig. \ref{fig:pcb_photos}(c) i). When improved electrical performance is required at higher frequencies, fencing vias and ground planes can be incorporated into the device board design using some or all of the grounded feed-throughs to ensure that both signal and device board ground are continuous.
\section{Crosstalk and Signal Fidelity Measurements}
Having described the layout and layer stack of the PCBs, we now present low temperature measurements characterizing the crosstalk and transmitted signal fidelity of the system.  The data shown here is taken at a temperature $T \sim\ $4 K in a high-frequency cryogenic probe-station\footnote[5]{LakeShore Cryotronics Corp. Model CRX-4K} with a calibrated vector network analyzer\footnote[6]{Agilent Corp. PNA5230C}. Using a dilution refrigerator we have verified that the $S$-parameters of the combined PCB/interposer system do not change when it is further cooled to below 20 milli-Kelvin. Data is shown over a bandwidth of 5 GHz for the rf lines and up to 12 GHz for the microwave lines.

Crosstalk performance was obtained via measurements of $S_{21}$ between unconnected lines with both the rf interposer and device board mounted\footnote[7]{In this geometry the dominant coupling mechanism will be capacitive and an open port represents the worst case performance. A shorted port would increase inductive coupling but is negligible in comparison.}. In order to set an upper bound on this value, nearest neighbour tracks which are most susceptible to crosstalk were used \cite{Blanvillian}. The transmission performance of the board was similarly obtained through an $S_{21}$ measurement with bond-wires used to connect adjacent bond-pads on top of the device board. We note that the use of bond-wires to bridge the transmission lines will degrade performance since they produce deviations in the characteristic impedance of the lines. In this sense the data reported is a lower bound for the board performance. Figure \ref{fig:measurements}(a) shows transmission and crosstalk data for nearest neighbour connections on the straight multiport connector lines passing through the interposer and device board (see Fig. \ref{fig:pcb_photos}(c) ii)). We observe that the transmission drops less than -3 dB for frequencies up to 5 GHz while crosstalk is kept below -40 dB to $\sim$ 4 GHz and below -20 dB to 5 GHz at both room temperature and 4 Kelvin. When cooled to cryogenic temperatures we observe a small decrease in resistive losses leading to an increase in both transmission and crosstalk. Corresponding data for the signal lines fed from the right-angle multiport connector is presented in Fig. \ref{fig:measurements}(b) where again we see that transmission loss remains below - 3 dB to 4.5 GHz while crosstalk is kept below -40 dB to ~2.5 GHz and below -20 dB over the entire measurement bandwidth. 

The performance of the microwave launchers is shown in Fig. \ref{fig:measurements}(c) up to a frequency of 12 GHz. Over this frequency range crosstalk remains below -40 dB to 4 GHz and below -20 dB to 12 GHz. 
In this case data is taken for just the signal board since current device boards are not designed for use at these frequencies. High frequency device boards can be implemented using low-loss dielectric substrates and by making use of coaxial geometries with appropriate grounds for signal lines feeding through the interposer. 
\section{Discussion and Conclusion}
As solid-state quantum circuits are scaled-up in the number and density of qubits, there is a corresponding need to engineer the electronic interconnect technology that connects these complex circuits to the outside world. Unlike industrial approaches to packaging integrated circuits where interconnect geometries are fixed, quantum devices are rapidly evolving as part of the research process. To facilitate the measurement of prototype quantum circuits we have designed and characterised a low-cost dual PCB solution which separates the signal interconnects and filtering from the bonded device itself using a high frequency interposer. This provides a modular framework with the high density of dc, rf, and microwave connections needed for quantum information processing. 
\section{Acknowledgements}
 We thank Xanthe Croot, Alice Mahoney, and John Hornibrook for technical assistance. This research was supported by the IARPA/MQCO program and the U. S. Army Research Office under Contract No. W911NF-11-1-0068 and the Australian Research Council Centre of Excellence Scheme (EQuS CE110001013).\\

* email: david.reilly@sydney.edu.au

\bibliographystyle{unsrtnat}

\end{document}